\pdfoutput=1
\RequirePackage{ifpdf}
\ifpdf 
\documentclass[pdftex]{sigma}
\else
\documentclass{sigma}
\fi

\usepackage[active]{srcltx}

\usepackage{dsfont}

\def\d{\delta}

\def\x{\xi}

\def\G{\Gamma}

\def\OO{\Omega}

\def\ca{{\cal A}}
\def\cb{{\cal B}}

\def\cg{{\cal G}}
\def\ch{{\cal H}}

\def\ck{{\cal K}}
\def\cl{{\cal L}}

\def\co{{\cal O}}
\def\cp{{\cal P}}

\def\cu{{\cal U}}

\newcommand{\pa}{\partial}

\newcommand{\bbN}{{\Bbb N}}

\newcommand{\bbC}{{\Bbb C}}

\newcommand{\bbR}{{\Bbb R}}

\newcommand{\bbZ}{{\Bbb Z}}

\newcommand{\Tr}{\operatorname{Tr}}

\fontfamily{yfrak}

\newcommand{\notD}{\mbox{$\not\hspace{-1.1mm}D$}}
\newcommand{\notA}{\mbox{$\not\hspace{-1.1mm}A$}}
\numberwithin{equation}{section}

\begin{document}

\allowdisplaybreaks

\renewcommand{\thefootnote}{$\star$}

\renewcommand{\PaperNumber}{018}

\FirstPageHeading

\ShortArticleName{Intersecting Quantum Gravity with Noncommutative Geometry -- a Review}

\ArticleName{Intersecting Quantum Gravity\\ with Noncommutative Geometry -- a Review\footnote{This
paper is a contribution to the Special Issue ``Loop Quantum Gravity and Cosmology''. The full collection is available at \href{http://www.emis.de/journals/SIGMA/LQGC.html}{http://www.emis.de/journals/SIGMA/LQGC.html}}}

\Author{Johannes AASTRUP~$^\dag$ and Jesper M{\o}ller GRIMSTRUP~$^\ddag$}

\AuthorNameForHeading{J.~Aastrup and J.M.~Grimstrup}

\Address{$^\dag$~Institut f\"ur Analysis, Leibniz Universit\"at Hannover, \\
\hphantom{$^\dag$}~Welfengarten 1, D-30167 Hannover, Germany}
\EmailD{\href{mailto:aastrup@math.uni-hannover.de}{aastrup@math.uni-hannover.de}}

\Address{$^\ddag$~Wildersgade 49b, 1408 Copenhagen, Denmark}
\EmailD{\href{mailto:grimstrup@nbi.dk}{grimstrup@nbi.dk}}

\ArticleDates{Received October 06, 2011, in f\/inal form March 16, 2012; Published online March 28, 2012}

\Abstract{We review applications of noncommutative geometry in canonical quantum gravi\-ty. First, we show that the framework of loop quantum gravity includes natural noncommutative structures which have, hitherto, not been explored. Next, we present the construction of a spectral triple over an algebra of holonomy loops. The spectral triple, which encodes the kinematics of quantum gravity, gives rise to a natural class of semiclassical states which entail emerging fermionic degrees of freedom. In the particular semiclassical approximation where all gravitational degrees of freedom are turned of\/f, a free fermionic quantum f\/ield theory emerges. We end the paper with an extended outlook section.}


\Classification{46L52; 46L87; 46L89; 58B34; 81R60; 81T75; 83C65; 70S15}

\renewcommand{\thefootnote}{\arabic{footnote}}
\setcounter{footnote}{0}

\section{Introduction}

The road to the resolution of the grand problem of theoretical physics -- the search for a unif\/ied theory of all fundamental forces~-- does not come with many road signs. The work by Connes and coworkers on the standard model of particle physics, where the standard model coupled to general relativity is reformulated as a single gravitational model written in the language of noncommutative geometry, appears to be such a road sign.

Noncommutative geometry is based on the insight, due to Connes, that the metric of a~compact spin manifold can be recovered from the Dirac operator together with its interaction with the algebra of smooth functions on the manifold~\cite{ConnesRecon}. In other words the metric is completely determined by the triple
\begin{gather*}
(C^{\infty}(M), L^2(M,S), \notD),
\end{gather*}
where $M$ is compact, oriented, smooth manifold, $S$ is a spin type bundle over $M$, and $\notD$ is a Dirac operator. This observation leads to a noncommutative generalization of Riemannian
geometries. Here the central objects are spectral triples $(A,H,D)$, where
$A$ is a not necessarily commutative algebra; $H$ a Hilbert space and $D$ an
unbounded self-adjoint operator called the Dirac operator. The triple is required
to satisfy some interplay relations between~$A$, $H$ and $D$ mimicking those
of $C^{\infty}(M)$, $L^2(M,S)$ and $\notD$. The choice of the Dirac operator~$D$ is strongly
restricted by these requirements.

The standard model of particle physics coupled to general relativity provides a key example of such a noncommutative geometry formulated in terms of a spectral triple \cite{Connes:1996gi}.
Here, the algebra is an almost commutative algebra
\[
A = C^{\infty}(M)\otimes A_F  ,
\]
where $A_F$ is the algebra $\mathbb{C}\oplus \mathbb{H} \oplus M_3(\mathbb{C})$. The corresponding Dirac operator
then consists of two parts,
$D = \notD + D_F$
where $D_F$ is
given by a matrix-valued function on the manifold $M$ that encodes the metrical
aspects of the states over the algebra $A_F$. It is a highly nontrivial and
very remarkable fact that the above mentioned requirements for Dirac operators
force $D_F$ to contain the non-Abelian gauge f\/ields of the standard model
and the Higgs-f\/ield together with their couplings to the elementary fermion f\/ields.

From the road sign, which we believe this formulation of the standard model is, we read of\/f three travel advices for the road
ahead:
\begin{enumerate}\itemsep=0pt
\item
It is a formulation of fundamental physics in terms of {\it pure geometry}. Thus, it suggests that one should look for a unif\/ied theory which is {\it gravitational} in its origin.
\item
The unifying principle in Connes formulation of the standard model hinges completely on the {\it noncommutativity} of the algebra of observables. Thus, it suggests that one should search for a suitable noncommutative algebra.
\end{enumerate}
We pick up the third travel advice from the fact that Connes work on the standard model coupled to general relativity is essentially classical. With its gravitational origin this is hardly a surprise: if the opposite was the case it would presumably involve quantum gravity and the problem of f\/inding a unif\/ied theory would be solved. This, however, suggests:
\begin{enumerate}\itemsep=0pt
\addtocounter{enumi}{2}
\item
That we look for a theory which is {\it quantum} in its origin.
\end{enumerate}
If we combine these three points we f\/ind that they suggest to look for a model of quantum gravity that involves an algebra of observables which is suf\/f\/iciently noncommutative and subsequently arrive at a principle of unif\/ication by applying the machinery of noncommutative geometry.
The aim of this review paper is to report on ef\/forts made in this direction. In particular we shall report on ef\/forts made to combine noncommutative geometry with canonical quantum gravity.

Loop quantum gravity \cite{Thiemann:2007zz} is an approach to {canonical quantum gravity} which is formulated in a language intriguingly similar to that of noncommutative geometry. It is based on a Hamiltonian formulation of gravity in terms of connections and triad f\/ields~-- the Ashtekar variables~\cite{Ashtekar:1987gu, Ashtekar:1986yd}. Thus, the conf\/iguration space is the space of Ashtekar connections and loop quantum gravity approaches the problem of quantizing these variables by choosing an algebra of observables over this conf\/iguration space. This algebra is generated by Wilson loops
\[
W(A,L) = {\rm Tr}\, \cp \exp \int_L A
\]
 and is constructed as an inductive limit of intermediate algebras associated to piece-wise analytic graphs. It is a key result, due to Ashtekar and Lewandowski \cite{Ashtekar:1993wf}, that the conf\/iguration space of connections is recovered in the spectrum of this algebra.

Thus, similar to noncommutative geometry, loop quantum gravity takes an algebra of functions as the primary object and recovers the underlying space~-- the conf\/iguration space of connections -- in a secondary step from the spectrum of the algebra. The algebra of Wilson loops is, however, commutative and does, therefore, a priori not prompt an application of noncommutative geometry. Yet, one immediately notices that the commutativity of this algebra is a contrived feature since an algebra generated by holonomy loops
\[
{\rm Hol}(A,L) = \cp \exp \int_L A
\]
 {\it is} noncommutative, and even more so, an algebra generated by parallel transports along open paths will be highly noncommutative. Thus, it is immediately clear that natural, noncommutative structures {\it do} exist within the basic setup of canonical quantum gravity formulated in terms of parallel transports. It is also clear that this noncommutativity is removed by hand in the loop quantum gravity approach.

These observations and considerations instigated a series of paper \cite{Aastrup:2011dt,Aastrup:2006ib,Aastrup:2005yk,AGN3,Aastrup:2009ux,AGN1,AGN2,Aastrup:2009dy,Aastrup:2010kb,Aastrup:2010ds,AGNP1}, which aims at investigating what additional structure this noncommutativity might entail, as well as exploring the possibility of employing the ideas and techniques of noncommutative geometry directly to this setup of canonical quantum gravity.

A f\/irst natural step towards this goal is to construct a Dirac type operator which interacts with a noncommutative algebra of holonomy loops.
In the papers \cite{AGN3, AGN1,AGN2} it was shown how such an operator is constructed as an inf\/inite sum of triad f\/ield operators -- the operators which quantize the Ashtekar triad f\/ields~-- which acts in a Hilbert space obtained as an inductive limit of intermediate Hilbert spaces associated to f\/inite graphs. This construction necessitates a couple of important changes made to the original approach of loop quantum gravity:
\begin{enumerate}\itemsep=0pt
\item
It is necessary to operate with a countable system of graphs.
\end{enumerate}
This is in marked contrast to loop quantum gravity which is build over the uncountable set of piece-wise analytic graphs. In~\cite{AGN2} it was shown that the central result due to Ashtekar and Lewandowski concerning the spectrum of the algebra of Wilson loops is also obtained with a~countable set of graphs. Thus, it is possible to separate the conf\/iguration space of connections with a countable set of graphs. Essentially, this means that the interaction between the Dirac type operator and the algebra of holonomy loops captures the kinematics of quantum gravity. Furthermore, a countable set of graphs entails a separable Hilbert space -- known as the kinematical Hilbert space -- in contrast to loop quantum gravity where the kinematical Hilbert space is nonseparable. This issue of countable versus uncountable is closely related to the question of whether or not and how one has an action of the dif\/feomorphism group.

Another important dif\/ference is that:
\begin{enumerate}\itemsep=0pt
\addtocounter{enumi}{1}
\item
The construction of the Dirac type operator necessitates additional structure in the form of an inf\/inite dimensional Clif\/ford bundle over the conf\/iguration space of connections.
\end{enumerate}
 This Clif\/ford bundle~-- which comes with an action of the CAR algebra~-- plays a key role in the subsequent results on semiclassical states and the emergence of fermionic degrees of freedom.
Indeed, in a second series of papers \cite{Aastrup:2011dt,Aastrup:2009dy,Aastrup:2010kb,Aastrup:2010ds, AGNP1} it was shown that a natural class of semiclassical states resides within this spectral triple construction. In a certain semiclassical approximation these states entail an inf\/inite system of fermions coupled to the ambient gravitational f\/ield over which the semiclassical approximation is performed. In one version of this construction these fermions come with an interaction which involves f\/lux tubes of the Ashtekar connection~-- in another version this interaction is absent. In any case, in the special limit where one turns of\/f all gravitational degrees of freedom -- that is, where one performs a semiclassical approximation around a f\/lat space-time geometry~-- a Fock space and a free fermionic quantum f\/ield theory emerge. Thus, a direct link between canonical quantum gravity and fermionic quantum f\/ield theory is established. These results seem to suggest that one should not attempt to quantize both gravitational and matter degrees of freedom simultaneously, but rather see the latter emerge in a semiclassical approximation of the former.

In this review paper we shall put special emphasis on open issues and point out where we believe this line of research is heading. In particular, we will end with an extensive outlook section. The paper is organized as follows: In Section~\ref{section2} we review the concepts and machinery of noncommutative geometry and topology. In Section~\ref{section3} we introduce canonical gravity, Ashtekar variables and the corresponding loop variables. In Section~\ref{section4} we show that in the quantization procedure of the loop variables one encounters natural noncommutative structures in the form of noncommutative algebraic structures over the conf\/iguration space of gravity. Section~\ref{section5} is devoted to the construction of a spectral triple over a particular algebra of holonomy loops and Section~\ref{section6} reviews the construction of semiclassical states. Section~\ref{section7} presents an extended outlook.

Let us end this introduction by noting that there exist in the literature also other lines of research which seek to combine elements of noncommutative geometry and loop quantum gravity. In~\cite{Denicola:2010ni} the aim is to encode information of the underlying topology in a spin-foam setting using monodromies and in~\cite{Lewandowski:2008ye} the loop quantum gravity setup is generalized to encompass also compact quantum groups.

\section{Noncommutative geometry} \label{section2}

In this section we will give a brief survey/introduction to noncommutative geometry in general. For more details we refer the reader to~\cite{ConnesBook} and \cite{ConnesMarcolliBook}. For background material on operator algebras we refer the reader to~\cite{KR1,KR2} and~\cite{Bratteli:1979tw,Bratteli:1996xq}.

In many situations in ordinary geometry, properties and quantities of a geometric space $X$ are  described dually via certain functions from $X$ to $\bbR$ or $\bbC$. Functions from $X$ to $\bbR$ or $\bbC$ come with a product, namely the pointwise product between two such functions. Due to the commutativity of $\bbR$ and $\bbC$ this product is commutative.

Noncommutative geometry is based on the fact that in many situations one considers objects with a noncommutative product, but which one would still like to treat with the methods and conceptual thinking of geometry. As an instance of this consider the Heisenberg relation
\[
[X,P]=i\hbar .
\]
One would of course like to think of $X$ and $P$ as functions on the quantized ``phase space'', however due to the Heisenberg relation this ``phase space" does not exist as a space.

We will in the following give some examples that illustrate geometrical aspects  often considered in noncommutative geometry. For now we want to outline some of the strengths  of noncommutative geometry:
\begin{enumerate}\itemsep=0pt
\item Many geometrical concepts and techniques can, when suitably adapted, be applied to objects beyond ordinary geometry.
\item  Noncommutative geometry comes with many tools (partly inspired by geometry, partly not), such as functional analysis, $K$-theory, homological algebra, Tomita--Takesaki theory, etc.
\item Constructions natural to noncommutative geometry capture structures of a unif\/ied framework, see for example the section on noncommutative quotients below.
\end{enumerate}

We have split the chapter into three subsections, dealing with topological aspects, measure theoretic aspects or metric aspects of noncommutative geometry. Readers mostly interested in the metric aspect of the noncommutative geometry and particle physics can skip the sections on noncommutative topology and von Neumann algebras.

\subsection{Noncommutative topology}\label{section2.1}

A possible noncommutative framework for topological spaces is the def\/inition of  $C^*$-algebras.

A $C^*$-algebra is an algebra $A$ over $\bbC$ with a norm $\| \cdot \|$, and an anti-linear involution $*$ such that $\|ab\|\leq \| a\|\| b\|$, $\|aa^* \|=\| a\|^2$ and $A$ is complete with respect to $\| \cdot \|$.

A fundamental theorem due to Gelfand--Naimark--Segal states that $C^*$-algebras can be equally well def\/ined  as norm closed $*$-invariant subalgebras of the algebra of bounded operators on some Hilbert space.

The following theorem states that the concept of $C^*$-algebras is the perfect generalization of  locally compact Hausdorf\/f-spaces.

\begin{theorem}[Gelfand--Naimark] \label{gn}
$C_0(X)$ is a commutative $C^*$-algebra.
Conversely, any commutative $C^*$-algebra has the form $C_0(X) $, where $X$ is a locally compact Hausdorff-space, and $C_0(X)$ denotes the algebra of continuous complex-valued functions on $X$ vanishing at infinity.
\end{theorem}

Given a commutative $C^*$-algebra, the space $X$,  for which $A=C_0(X)$ is given by
\begin{gather*}
X= \{ \gamma :A \to \bbC \,| \, \gamma \hbox{ nontrivial }C^*\hbox{-homomorphism} \}
\end{gather*}
equipped with the pointwise topology. $X$ is also called the spectrum of $A$.

More generally, given a not necessarily commutative $C^*$-algebra $A$, the spectrum is def\/ined as the set of all irreducible representations of $A$ on Hilbert spaces modulo unitary equivalence.

{\bf The rotation algebras.}
Because of Theorem~\ref{gn} the 2-dimensional torus can be seen as the universal $C^*$-algebra  generated by two commuting unitaries, i.e.\
$U_1$, $U_2$ satisfying
\begin{gather*}
U_1U_2=U_2U_1, \qquad U_1U_1^*=U_1^*U_1=U_2U_2^*=U_2^*U_2=1.
 \end{gather*}
For a given $\theta \in \bbR$ the  rotation algebra $A_\theta$ is given as the universal $C^*$-algebra generated by two unitaries satisfying
\[
U_1U_2=e^{2 \pi i\theta }U_2U_1, \qquad U_1U_1^*=U_1^*U_1=U_2U_2^*=U_2^*U_2=1.
\]

The structure (in particular the $K$-theory) of this algebra plays an important role in the understanding of the integer-valued quantum Hall ef\/fect, see for example \cite{Bellisard2,Bellisard1, ConnesBook}.

{\bf Duals of groups.}
Another example, where noncommutativity provides an advantage, is in describing the duals of groups. Let  for simplicity $G$ be a discrete group (what follows also works for $G$ locally compact, or even locally compact groupoid with a left Haar-system, but one has to be more careful with def\/initions). Then $G$ acts naturally on $L^2(G)$ via
\[
 U_g(\xi) (h)=\xi \big(hg^{-1}\big),\qquad \xi \in L^2(G).
 \]
Def\/ine $C_r^*(G)$ to be the closure of all linear combinations of the $U_g$'s in the norm topology of $\cb (L^2(G) )$, the bounded operators on $\cb (L^2(G) )$. It can be shown:
\begin{theorem}
When $G$ is commutative, then
\[
 C^*_r(G)\simeq C_0(\hat{G}),
 \]
where $\hat{G}$ is the dual of $G$, i.e.
\[
\hat{G}=\{ \rho :G\to U(1) \,|\, \rho \hbox{ group homomorphism}\} .
\]
\end{theorem}
For example if $G=\bbZ$, $\hat{G}=U(1)$ and the isomorphism is the Fourier transform. If $G=\bbR$, $\hat{G}=\bbR$ and the isomorphism is the Fourier transform, i.e.\ the Fourier transform maps convolution product to pointwise product.

When $G$ is non-Abelian, $\hat{G}$  does not contain much information. However $C^*_r(G)$ continues to make sense, and contains a lot of information about $G$. We can therefore consider $C^*_r(G)$ as the replacement of $\hat{G}$

In fact computations of the $K$-theory of $C^*_r(G)$, also known as the Baum--Connes conjecture, has led to deep and new insights to topology and group theory, see for example~\cite{Valette}.

{\bf Noncommutative quotients.}
We will start with the example of two points $\{ a, b\}$ with the relation $a \sim b$. In ordinary topology the quotient $\{ a, b\} / {\sim} $ is just the one point set. However in the noncommutative setting we consider f\/irst the algebra of functions over $\{ a, b\}$, namely $\bbC\oplus \bbC$. Given a function $f$ on $\{ a,b\}$ we will write it $(f_a,f_b)$. Now instead of identifying the two copies of $\bbC$, we embed them into the larger algebra $B$ in which we ``identify'' the two copies of $\bbC$ by inserting partial unitaries $U_{ab}$ and $U_{ba}$ mapping between the two copies of $\bbC$, i.e.
\[
 U_{ba} (f_a,f_b)U_{ab}=(0,f_a)\qquad \mbox{and} \qquad U_{ab} (f_a , f_b)U_{ba}=(f_b,0).
 \]

{\sloppy  We then def\/ine functions on the noncommutative quotient, which we will denote $ C_{nc}(\{ a, b\} / {\sim} )$,  to be the algebra generated by $\bbC\oplus \bbC$ and $U_{ab}$, $U_{ba}$. It is immediate that via the identif\/ication
\[(f_a,f_b, x_{ab}U_{ab}, x_{ba}U_{ba}) \to   \begin{pmatrix}
f_a & x_{ba} \\x_{ab} & f_b
\end{pmatrix}, \qquad x_{ab},x_{ba}\in \bbC
 \]
we get an isomorphism with the two-by-two matrices. We therefore see that the noncommutative quotient has the same representation theory as the commutative one, in particular the spectrum of $ C_{nc}(\{ a, b\} / {\sim} )$ is just a single point. However the noncommutative quotient has a richer structure. For example the noncommutative quotient has states $| a \rangle$, $|b\rangle $, i.e.
\[
\langle a |(f_a,f_b, x_{ab}U_{ab}, x_{ba}U_{ba}) | a \rangle =f_a, \qquad \langle b |(f_a,f_b, x_{ab}U_{ab}, x_{ba}U_{ba}) | b \rangle =f_b .
\]

}

For a more interesting example we can consider the circle $S^1$ (which we identify with $\{ e^{2\pi i \varphi} \}$, $\varphi \in [0,1]$) with the action $\alpha_\theta$ of $\bbZ$ given by
\[
 \alpha_\theta (n)( e^{2\pi i \varphi } ) =e^{2\pi i (\varphi -n\theta)}.
 \]
It is not hard to see that when $\theta $ is irrational the  quotient $S^1 / \bbZ $ becomes $[0,1] /( \mathbb{Q} \cup [0,1])$ with the dif\/fuse topology, i.e. the topology with only two open sets. The quotient therefore carries no information of the original situation.
If we instead form the noncommutative quotient using the construction from above, we enlarge the algebra $C(S^1)$ with partial unitaries  $U_n$, $n\in \bbN$, such that
\[
U_n fU_{-n} = \alpha^*_\theta (n) (f),
\]
 $\alpha^*_\theta$ denoting the action on $C(S^1)$ induced by $\alpha_\theta$.  By applying $f=1$ we see that $U_n$ are unitaries with $U_n^*=U_{-n}$. It is furthermore natural to impose that the action of $U_n$ on $C(S^1)$ determines~$U_n$. We therefore get, since $\alpha_\theta$ is a group homomorphism that $U_nU_m=U_{n+m}$, $n,m \in \bbZ$, i.e.~$\bbZ \ni n \to U_n$ is a group homomorphism. We will denote the algebra generated by $C(S^1)$ and the $U_n$'s by  $C(S^1)\times_{\alpha_\theta} \bbZ$. Since $n\to U_n$ is a group homomorphism, and since $C(S^1)$ is generated by one unitary $V=e^{2\pi i \varphi}$, we see that $C(S^1)\times_{\alpha_\theta} \bbZ$ is generated by two unitaries~$V$,~$U_1$ satisfying
 \[
  U_1 V U_1^* =e^{2\pi i \theta} V ,
  \]
 i.e.~$C(S^1)\times_{\alpha_\theta} \bbZ$ is just $A_\theta$.
We have therefore obtained an object with a much richer and interesting structure than the ordinary quotient.

This construction works more generally for a locally compact group $G$ acting on a space $X$ or a $C^*$-algebra. The special case when $G$ is acting on a point leads to the group $C^*$-algebra.

For more justif\/ication and a longer list of interesting example of noncommutative quotients, see Chapter 2 in~\cite{ConnesBook}.

\subsection[von Neumann algebras and Tomita-Takesaki theory]{von Neumann algebras and Tomita--Takesaki theory}\label{section2.2}

The natural setting for noncommutative measure theory is von Neumann algebras. A von Neumann algebra is a subalgebra~$M$ of the bounded operators on a Hilbert space $\ch$, which is closed under  adjoints and satisfying
\[
(M')'=M  ,
\]
where
\[
M'=\{ a \in \cb (\ch )\, |\, ab=ba \ \hbox{for all} \ b\in M\}.
\]
The famous bicommutant theorem of von Neumann states that this property is equivalent to~$M$ acting nondegenerate on $\ch$ and being closed in the weak operator topology on $\cb (\ch )$.

Commutative von Neumann algebras admitting faithful representations on a separable Hilbert space have the form of bounded measurable functions on a second countable locally compact Hausdorf\/f space equipped with a probability measure. This ensures that von Neumann algebras are the natural generalization of measure spaces.

The probably most surprising and interesting feature of von Neumann algebras, is that they have a canonical time f\/low. More precisely a von Neumann algebra admits a one parameter group of automorphisms, which is unique up to inner equivalence.
The way it appears is the following (see \cite{Bratteli:1979tw, take} for details):

Let us suppose that $M$ is represented on $\ch$, and that this representation admits a  separating and cyclic vector $\xi$,
i.e. \[
m\xi =0 \ \Leftrightarrow \ m=0 \quad \hbox{for all} \ m \in M,
\]
and the closure of $\{ m\xi \, | \, m\in M\}$ is~$\ch$.
 There is an anti-linear usually unbounded operator~$S$ on~$\ch$ def\/ined via
 \[
 S(m\xi )=m^*\xi .
 \]
 This operator admits a polar decomposition
 \[
 S=J\Delta^{\frac12},
  \]
 where $J$ is an anti-unitary and $\Delta$ a selfadjoint positive operator. The time f\/low is the given by
 \[
  \alpha_t^\xi(m)=\Delta^{-it} m \Delta^{it} , \qquad m \in M.
 \]
 This group of automorphisms is dependent of $\xi$. The cocycle Radon--Nikodym theorem by Connes, see \cite{ConnesNeumann}, ensures that given another cyclic separating vector $\eta$  there exist a one parameter family of unitaries $U_t$ in $M$ satisfying
 \[
 \alpha_t^\eta(m)=U_t\alpha_t^\xi(m)U_t^*, \qquad \hbox{for all} \quad t\in \bbR \quad \hbox{and} \quad m\in M,
 \]
and
\[
 U_{t_1+t_2} = U_{t_1} \alpha^\xi_{t_1} (U_{t_2} ) , \qquad \hbox{for all}\quad t_1,t_2 \in \bbR.
 \]
This in  particular means that up to inner unitary equivalence, the time f\/low is independent of the representation and the cyclic separating vector.

This result has the potential to become a major principle for def\/ining time in physical theories. Finding the algebra of observables automatically gives, up to inner automorphisms, a canonical  notion of time. For more details on applications to relativistic quantum f\/ield theory see~\cite{Borchers:2000pv, Haag:1992hx} and see~\cite{ConnesRovelli} for the notion of thermal time.

It is important to note that this time-f\/low is a purely noncommutative concept, since it vanishes for commutative von Neumann algebras.

\subsection{Noncommutative Riemannian geometry }\label{section2.3}

So far we have only been dealing with noncommutative topology and measure theory. What is missing is the metric structure. The crucial observation by Alain Connes is, that given a~compact manifold\footnote{If the manifold is nonconnected the geodesic distance can assume inf\/inite values.} $M$ with a metric $g$, the geodesic distance $d_g$ of $g$, and thereby also $g$ itself, can be recovered by the formula
\begin{gather}
d_g(x,y)=\sup \{ |f(x)-f(y)| \, |\, f\in C^\infty (M)\ \hbox{with} \ \| [ \notD,f]\| \leq 1   \}   , \label{connesdist}
\end{gather}
where $ \notD$ is a Dirac type operator associated to $g$ acting in $L^2 (M,S)$, $S$ is some spinor bundle, $\|[ \notD,f]\|$ the operator norm of $[ \notD,f]$ as operator in $L^2 (M,S)$. Therefore to specify a metric, one can equally well specify the triple
\[
 \big(C^\infty (M), L^2(M,S),  \notD\big).
 \]

This observation leads to the def\/inition:
\begin{definition} \label{spectrip}
A spectral triple $(\cb , \ch ,D)$ consists of a unital $*$-algebra   $\cb$ (not necessary commutative), a separable Hilbert space $\ch$, a unital $*$-representation
\[
\pi : \ \cb \to \cb (\ch)
\]
and a  self-adjoint operator $D$ (not necessary bounded) acting on $\ch$ satisfying
\begin{enumerate}\itemsep=0pt
\item $\frac{1}{1+D^2} \in \ck (\ch)$; \label{kompakt}
\item $[D,\pi (b)]\in \cb(\ch)$ for all $b\in \cb  $. \label{diffe}
\end{enumerate}
where $\ck(\ch)$ are the compact operators.
\end{definition}

This def\/inition is the replacement for metric spaces in the noncommutative setting.

 Note that $(C^\infty (M), L^2(M,S),  \notD)$ fulf\/ils (\ref{kompakt}) and~(\ref{diffe}).  Property~$1$ ref\/lects the fact that the absolute values of the eigenvalues of $ \notD$ converges to inf\/inity and that each eigenvalue only have f\/inite degeneracy. Property $2$ ref\/lects the fact that the functions in $C^\infty (M)$ are dif\/ferentiable.

The change of viewpoint of going from a metric to the Dirac operator can be interpreted physically in the following way: Instead of measuring distances directly in space, one measures ``frequencies'', i.e.\ eigenvalues of the Dirac operator and its interaction with the observables on the manifold (smooth functions on~$M$).

Def\/inition~\ref{spectrip} is insuf\/f\/icient as a def\/inition of noncommutative Riemannian geometry. In fact it can be shown, that all compact metric spaces f\/it into Def\/inition~\ref{spectrip}, see~\cite{ChrisIvan}. Therefore to pinpoint a def\/inition of a noncommutative Riemmanian  manifold one needs to add more axioms to those of a spectral triple.

In \cite{ConnesRecon} it was shown, that given a commutative spectral triple satisfying the extra axioms specif\/ied in~\cite{ConnesRecon} it is automatically an oriented compact manifold.
We will not give the details here, but refer to~\cite{ConnesRecon}. For a set of axioms for noncommutative oriented Riemmanian geometry see~\cite{LRV},  for the original axioms of noncommutative spin manifolds see~\cite{Connes:1996gi} and for a generali\-za\-tion to almost commutative geometries and weakened the orientability hypothesis, see~\cite{cacic}.
There is however one important aspect we want to mention here. In noncommutative spin geo\-met\-ry there is an extra ingredient, the real structure~$J$, which plays an important role. For a~four-dimensional spin manifold~$J$ is the charge conjugation operator. In general $J$ is required to be an anti-linear operator on~$\ch$ with the property that~$Ja^*J^{-1}$ gives a right action of~$A$ on~$\ch$, and satisfying some additional axioms.

\begin{example}[The two point example] 
The algebra for the two point space $\{ a,b \} $ is $\bbC\oplus \bbC $. The Hilbert space for the spectral triple will also be  $\bbC\oplus \bbC$. We choose the Dirac operator to be
\[
 D =   \begin{pmatrix}
0 & \lambda \\ \lambda & 0
\end{pmatrix} , \qquad \lambda \in  \bbR \setminus \{ 0\}.
\]
This clearly is a spectral triple. If we use the distance formula~(\ref{connesdist}) a small computation gives $d(a,b)= | \lambda |^{-1}$.
\end{example}

\begin{example}[Matrix valued functions] \label{matrix}
We consider the algebra $A=C^\infty (M)\otimes M_n$, where $M$ is a compact manifold.  We represent $A$ on $\ch =L^2(M,M_n\otimes S )$, where $S$ is a spinor bundle, in the obvious way. The Dirac type operator~$ \notD$ acting in $S$ acts naturally on~$\ch$. From the commutative case it follows that $(A,\ch ,  \notD) $ is a spectral triple.

This triple admits a real structure $J$. For $\xi \in \ch$ with $\xi (m)= c \otimes s\in M_n\otimes S$ it is given by:
\[
 (J\xi)(m)=c^*\otimes J_S (s) ,
 \]
where $J_S$ is the real structure on $S$ (in 4 dimensions the charge conjugation operator).
It follows that $Ja^*J^{-1}$ is simply pointwise multiplication on the matrix factor $M_n$ in $\ch$ from the right.
\end{example}

{\bf Inner f\/luctuations and one forms.}
For the triple $(C^\infty (M), L^2(M,S),  \notD)$ one can identify one forms on $M$ with elements of the form
\[
 f_i[ \notD,g_i] , \qquad f_i,g_i \in C^\infty (M).
 \]
This comes about since $[ \notD,f]=df$, where $d$ is the exterior derivative. For a general spectral triple $(\cb,\ch, D)$ it is therefore natural to call elements of the form 	
\[
a_i[D,b_i], \qquad a_i,b_i \in \cb
\]
for noncommutative one forms.

We let $\cu$ be the group of unitary elements in $\cb$. Given $u\in \cu$, $D$ is in general not invariant under conjugation with $u$, but transforms according to
\[
 D \to D+u[D,u^*] .
 \]
It is therefore natural to propose an invariant form of~$D$ as
\[
 D_A=D+A,
\]
where $A$ is a general self-adjoint one-form, i.e.\ $A=A^*$. Under the  action of~$\cu$ we see that $uD_Au^*=D_{G(u)(A)}$, where
\[
G(u)(A)=uAu^*+u[D,u^*].
\]

Note that for $(C^\infty (M), L^2(M,S),  \notD)$, $M$ Minkowski spacetime, this is the transition
\[
 i \overline{\psi} \gamma^\mu \partial_\mu \psi \to  i \overline{\psi} \gamma^\mu (\partial_\mu + eA_\mu ) \psi
\]
in order to maintain $U(1)$-gauge invariance.
Due to the identif\/ication $[ \notD,f]=df$, $f\in C^\infty (M)$ the noncommutative one-forms correspond to the $U(1)$-gauge potentials.
Furthermore $G(u)$  is the gauge transformation induced by $u$.

For the case of matrix valued functions on a manifold the noncommutative one-forms can be identif\/ied with $U(n)$-gauge f\/ields, and the action of $\cu (C^\infty (M)\otimes M_n)$ on the $U(n)$-gauge f\/ields is the $U(n)$-gauge action.

Therefore the invariant operator $D_A$ is the framework for the gauge sector for spectral triples.

 In the presence of a real structure $J$ one requires invariance under the adjoint action $\xi \to u\xi u^*$ rather than $\xi \to u\xi$.
 In this case $D$ transforms according to
 \[
  D \to D+u[D,u^*]+\epsilon J u[D,u^*]J^* ,
 \]
 where $\epsilon$ is a certain sign depending on the real-dimension of $(\cb , \ch, D, J)$, and the invariant operator is
\[
D_A=D+A+\epsilon JAJ^{-1} ,
\]
 $A$ self-adjoint one-form.

The computation of the noncommutative one-forms in Example~\ref{matrix} in the presence of the real structure can be found in~\cite{ConnesMarcolliBook}. The one-forms can be identif\/ied with $SU(n)$-gauge f\/ields, and the action of $\cu (C^\infty (M)\otimes M_n)$ on the $SU(N)$-gauge f\/ields descends to a $PU(N)$ gauge action. Note in particular that for commutative case there are no gauge f\/ields, i.e.\ the gauge sector is a~purely noncommutative ef\/fect.

In the two point example a general noncommutative one form has the form
\[   \begin{pmatrix}
0 &   \Phi \\
   \bar{\Phi} & 0
\end{pmatrix}, \qquad \Phi \in \bbC.
\]
When suitably combined with the manifold case, the $\Phi$ will become the Higgs gauge boson, see for example Chapter~6.3 in~\cite{ConnesBook}.

\subsubsection{The standard model}
Some of the surprising outcomes of noncommutative geometry is the natural incorporation of the standard model coupled to gravity into the framework, and in particular the severe restrictions this puts on other possible models in high energy physics. Since the details of this are very subtle and elaborate,  we will omit most of the details here, and refer the reader to \cite{Chamseddine:2006ep}.

The basics of the construction is to combine the commutative spectral triple
\[
(C^\infty (M), L^2(M,S),  \notD ),
 \]
 where $M$ is a 4-dimensional manifold and a f\/inite dimensional triple $(\ca_F , \ch_F, D_F)$, which is a~variant on the two-point triple described above, by tensoring them, i.e.
\[
(C^\infty (M)\otimes \ca_F , L^2(M,S)\otimes \ch_F ,  \notD\otimes 1 +\gamma_5\otimes D_F) .
\]
Of course the exact structure of $(\ca_F , \ch_F, D_F)$ is to a large degree constructed from the standard model, and the Hilbert space $\ch_F$ labels the fermionic content of the standard model. Elements $\psi \in  L^2(M,S)\otimes \ch_F$ describe the fermionic f\/ields.

Given this triple the noncommutative dif\/ferential forms generate the gauge sector
of the standard model and the action of the standard model minimally coupled to the Euclidean background given by $\notD$ is given by
\[
\cl (A,\psi ) =\Tr \phi\left( \frac{D_A^2}{\Lambda^2} \right) +\langle J\psi | D\psi \rangle ,
\]
where $\phi$ is a suitable function and $\Lambda$ is a cutof\/f.

Some of the appealing features of this setting of the standard model are the following
\begin{itemize}\itemsep=0pt
\item The standard model coupled to gravity is treated in a unifying framework.
\item The Higgs boson is an output and not an input.
\item There are strong constraints on the particle physics models which f\/it into this framework. This makes it a very delicate issue to extend the noncommutative geometry framework to a broad range of models.
\end{itemize}
Some of the limitations at the present moment are
\begin{itemize}\itemsep=0pt
\item The formulation gives the Lagrangian, and therefore a quantization scheme has to be applied afterwards.
\item The formulation only works in Euclidean signature.
\end{itemize}

\section{Connection formalism of gravity}\label{section3}

In this section we brief\/ly recall the formulation of canonical gravity in terms of connection and loop variables (for details see \cite{AL1,Dona:2010hm,Rovelli:2004tv,Sahlmann:2010zf}). This formulation permits a quantization procedure based on algebras generated by parallel transports and, subsequently, the construction of a~spectral triple over such an algebra.

First assume that space-time $M$ is globally hyperbolic. Then $M$ can be foliated as
\[
M=\Sigma \times \bbR,
\]
where $\Sigma$ is a three-dimensional hyper surface. We will assume that $\Sigma$ is oriented and compact.

The f\/ields, known as the Ashtekar variables \cite{Ashtekar:1987gu, Ashtekar:1986yd}, in which we will describe gravity are\footnote{Here we introduce the real Ashtekar connections, which corresponds either to the Eucledian setting or to a~formulation where the constraints, see (\ref{constraints}), are more involved. The original Ashtekar connection is a complexif\/ied $SU(2)$ connection.}
\begin{itemize}\itemsep=0pt
\item $SU(2)$-connections in the trivial bundle over $\Sigma$. These will be denoted $A_i^a$, where $a$ is the $\mathfrak{su}(2)$-index.
\item $\mathfrak{su}(2)$-valued vector densities on $\Sigma$. We will adopt the notation $E_a^i$.
\end{itemize}
On the space of f\/ield conf\/igurations, which we denote $\mathcal{P}$, there is a Poisson bracket expressed in local coordinates by
\begin{gather}
\{ A_i^a (x), E_b^j(y) \} =\delta_i^j \delta_b^a \delta (x,y),
\label{poissonbracket}
\end{gather}
where $\delta (x,y)$ is the delta function on $\Sigma$. The rest of the brackets are zero.
These f\/ields are subjected to constraints (Euclidean signature) given by
\begin{gather}
 \epsilon_c^{ab} E^i_a E^j_b F_{ij}^c = 0,\qquad
   E^j_a F^a_{ij} = 0,\qquad
  (\partial_i E^i_a+\epsilon_{ab}^cA_i^b E^i_c) = 0.
\label{constraints}
\end{gather}
Here $F$ is the f\/ield strength tensor, of the connection~$A$. The f\/irst constraint is the Hamilton constraint, the second is the dif\/feomorphism constraint and the third is the Gauss constraint.

These f\/ield conf\/igurations together with the constraints constitute an equivalent formulation of General Relativity without matter.

\subsection{Reformulation in terms of holonomy and f\/luxes}\label{section3.1}

The formulation of gravity in terms of connection variables permits a reformulation of the Poisson bracket in terms of holonomies and f\/luxes. For a given path $p$ in $\Sigma$ the holonomy function is simply the parallel transport along the the path, i.e.
\[
\mathcal{P}\ni (A,E) \to {\rm Hol} (p,A)\in G ,
\]
where we take $G$ to equal $SU(2)$. Given an oriented surface $S$ in $\Sigma$ the associated f\/lux function is given by
\[
\mathcal{P}\ni (A,E)\to \int_S \epsilon_{ijk} E_a^idx^jdx^k.
\]
The holonomy function for a path will also be denoted with $h_p$ and the f\/lux function will be denoted by $E^S_a$.

Let $p$ be a path and $S$ be an oriented surface in $\Sigma$ and assume that $p$
ends in $S$ and has exactly one intersection point with $S$. The Poisson bracket~(\ref{poissonbracket}) in this case becomes
\begin{gather}
\{ h_p, E^S_a \}(A,E)=\pm \frac{1}{2} h_p(A)\sigma_a,
\label{poisson1}
\end{gather}
where $\sigma_a$ is the Pauli matrix with index $a$. The sign in (\ref{poisson1}) is negative if the orientation of~$p$ and~$S$ is the same as the orientation of $\Sigma$, and positive if not. If~$p$ instead starts on~$S$ one gets the reverse sign convention.

\section[$C^*$-algebras of parallel transports]{$\boldsymbol{C^*}$-algebras of parallel transports}
\label{section4}

With the classical setup in place the next step is to settle on a quantization strategy in which the parallel transport- and f\/lux variables and their Poisson structure are represented as operators in a Hilbert space. Here, we shall in fact attempt to do something more: inspired by the classical setup we wish to identify canonical structures at a quantum level which, in a secondary step, entail known physical structures in a semiclassical analysis. The identif\/ication of a spectral triple build from these variables is such a structure.

Thus, we start by identifying which type of algebras can be constructed based on the classical setup.

\subsection[Three types of $C^*$-algebras]{Three types of $\boldsymbol{C^*}$-algebras}\label{section4.1}

We f\/irst outline which type of $C^*$-algebras of quantized observables we can construct from the classical loop and f\/lux variables.
We shall here only consider the parallel transport variables since the operators which quantize the f\/lux variables are easily introduced once a suitable algebra of parallel transports is def\/ined, see Section~\ref{section5.1}.

Notice f\/irst that a parallel transport along a path $p$ is a map
\[
h_p: \ \ca \rightarrow M_k,
\]
where $M_k$ are $k$ by $k$ matrices corresponding to a matrix representation of the gauge group and~$\ca$ denotes the conf\/iguration space of connections\footnote{For simplicity we assume that the connections in $\ca$ are in a trivial bundle.}.
Two parallel transports can be multiplied by composition
\[
h_{p_1}\cdot h_{p_2} =
\begin{cases}
h_{p_1\cdot p_2}, \ & \mbox{if} \  e(p_1)= s(p_2),\\
0, & \mbox{otherwise},
\end{cases}
\]
where $e(p)$, $s(p)$ denote the end point and start point of a path~$p$. Further, there is a natural $*$-operation given by the inversion of the direction of~$p$.

Basically, there appear to be three dif\/ferent ways in which one can construct a $*$-algebra generated by parallel transports:
\begin{enumerate}\itemsep=0pt
\item
{\bf Wilson loops:} one may consider the algebra generated by traced holonomy loops, i.e.\ Wilson loops. This construction requires a choice of {\it base point} in order to have a product between holonomy loops. Thus, the loops considered are based loops that start and end at the base point. The choice of Wilson loops is easily justif\/ied since these are gauge invariant objects. Even more so, due to the trace the dependency on base point vanishes since a change of base point amounts to a conjugation with a parallel transport between the old and the new base point, and such a conjugation vanishes under a trace. An algebra generated by Wilson loops is commutative.
\item
{\bf Holonomy loops:} one may alternatively consider an algebra generated by un-traced holonomies. Such an algebra will be noncommutative, although the noncommutativity of this algebra will simply be the noncommutativity of the gauge group. Furthermore, an algebra generated by holonomy loops will, a priori, be base point dependent. This dependency can, however, be shown to vanish on physical semiclassical states (see \cite{Aastrup:2010ds} and Section~\ref{section6.2}).
\item
{\bf Parallel transports:} f\/inally, one may consider an algebra generated by parallel transports along open paths with a groupoid structure. Again, this will be a noncommutative algebra\footnote{For this to generate an algebra and not an algebroid we def\/ine the product between two paths which do not meet to be zero.}, and in this case the noncommutativity is both due to the gauge group as well as the noncommutative structure of the groupoid of paths. For instance, if $e(p_1)=s(p_2)$ and $s(p_1)\not= e(p_2)$ then
\[
 h_{p_1}\cdot h_{p_2} - h_{p_2}\cdot h_{p_1} =h_{p_1}\cdot h_{p_2} \not=0  .
\]
An algebra generated by parallel transports will not depend on any base point.
\end{enumerate}
The f\/irst approach has been studied extensively in the loop quantum gravity literature. The second approach is the subject of this review and has been studied in the papers~\cite{Aastrup:2011dt,Aastrup:2006ib,Aastrup:2005yk,AGN3,Aastrup:2009ux,AGN1,AGN2,Aastrup:2009dy,Aastrup:2010kb,Aastrup:2010ds,AGNP1}. The third approach has, to our knowledge, not been studied in the literature (although the idea of using groupoids in the context of quantum gravity was proposed in~\cite{AGN1, Martins:2010xh}). We shall comment on the third possibility in Section~\ref{section7}.

Concretely, these algebras are constructed via inductive limits over intermediate algebras associated to f\/inite graphs. Thus, one chooses an inf\/inite set $\OO=\{\G_n\}$ of directed, f\/inite graphs with directed edges. Here  the index $n$ need not be countable. Here, directed means that $\OO$ is a~collection of graphs with the requirement that for any two graphs $\G_1$, $\G_2$ in $\OO$ there will exist a~third graph $\G_3$ in $\OO$ which includes $\G_1$ and $\G_2$ as subgraphs (for details see~\cite{AGN2}). For each graph~$\G$ one introduces a f\/inite dimensional space
\begin{gather}
\ca_\G = G^{n(\G)},
\label{intermediate}
\end{gather}
where $n(\G)$ is the number of edges in $\G$ and where a copy of the gauge group~$G$  is assigned to each edge in~$\G$.
A smooth connection~$A$ gives rise to a point in~$\ca_{\G}$ via
\[
\ca \ni \nabla \to  ({\rm Hol} ( A, e_i))_{e_i\ {\rm edge \ in}\ \Gamma},
\]
and therefore one should interpret the space~$\ca_{\G}$ as an approximation of the space~$\ca$, restricted to the graph~$\G$.

There are canonical maps
\[
P_{n+1,n} : \ \ca_{\G_{n+1}} \to \ca_{\G_n},
\]
which simply consist of multiplying elements in $G$ attached to an edge in $\Gamma_n$ which gets subdivided in $\Gamma_{n+1}$. Def\/ine
\[
\overline{\ca}=\lim_n \ca_{\G_n},
\]
where the limit on the right hand side is the projective limit. Since each $\ca_{\G_n}$ is a compact topological Hausdorf\/f space, $\overline{\ca}$ is a compact Haussdorf space. It is easy to see that the maps $\ca \to \ca_{\G_n}$ induce a map
\begin{gather}
\ca \to \overline{\ca} .
\label{mapmap}
\end{gather}
This map is a dense embedding when $\OO$ satisf\/ies certain requirements spelled out in~\cite{AGN2}. In particular, the set of nested, cubic lattices entails a~dense embedding.

We shall now restrict ourselves to the case of based loops.
Thus, we choose a f\/ixed base point~$x_0$ and consider loops~$L$ running in~$\G$ which start and end in~$x_0$. There is a natural product between such loops, simply by composition at the base point, and an involution by reversal of the loop. A~loop~$L$ in~$\G_n$ gives rise to a function~$h_L$ on~$\ca_{\G_n}$
\[
 \ca_{\G_n}  \ni \nabla  \rightarrow h_L(\nabla) \in      M_m(\mathbb{C}),
\]
where $m$ is the size of a matrix representation of the group~$G$ and where~$h_L(\nabla)$ is the composition of the various copies of~$G$ corresponding to the edges which the loop runs through.
The algebras~$\cb_{\G_n}$ are generated by sums
\[
a = \sum_i a_i h_{L_i} ,\qquad a_i\in\mathbb{C} , \qquad L_i\in\G_n ,
\]
with the obvious product and involution. There is a natural norm
\begin{gather}
\vert\vert a \vert \vert = \sup_{\nabla\in\ca_{\G_n}}  \bigg\vert\bigg\vert \sum_i a_i   h_{L_i}(\nabla)  \bigg\vert\bigg\vert_G  ,
\label{norm}
\end{gather}
where $ \vert  \vert\cdot  \vert \vert_G$ on the r.h.s.\ is the matrix norm. Finally, the limit $*$-algebra $\cb$ is obtained as an inductive limit
\[
\cb = \lim_n \cb_{\G_n} .
\]

In general, the construction of all three types of algebras depends heavily on which type of graphs one chooses. In particular, the choice whether one works with a countable or uncountable set of graphs is fundamental. The set of piece-wise analytic graphs used in loop quantum gravity is an uncountable set, a feature that makes the Hilbert space, which carries a representation of the corresponding algebra, nonseparable. The construction which we review in this paper is based on a countable set of cubic lattices, a setup which entails a separable Hilbert space. Whichever approach is the right one is yet to be determined. It is, however, clear that this issue is closely related to the question whether or not one has an action of the dif\/feomorphism group, and how this action is implemented. We shall comment on this in Section~\ref{section7.3}.

Let us end this section by pointing out that the key requirement when choosing a set of graphs is to ensure that one captures the full information of the underlying conf\/iguration space~$\ca$ of connections. This space is the conf\/iguration space of quantum gravity and it is the tangent space hereof~-- def\/ined in whichever way~-- which contains the quantization of gravity. In other words, the full conf\/iguration space of connections, or possible a gauge invariant section hereof, must be fully contained in the spectrum of the algebra. This requirement is related to the fact that the map~(\ref{mapmap}) is a dense embedding.

\section{A spectral triple of holonomy loops}\label{section5}

\subsection{The Hilbert space and the f\/lux operators}
\label{section5.1}

\begin{figure}[t]
\centering
\includegraphics{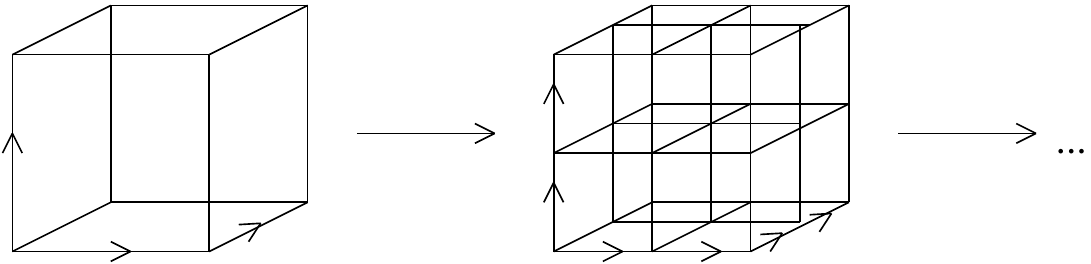}
\caption{One subdivision of a cubic lattice.} \label{ronnie}
\end{figure}

We are now ready to construct the Hilbert space carrying a representation of an algebra generated by holonomy loops together with the f\/lux operators. For the remainder of this paper we shall restrict ourselves to a set of cubic lattices.
Thus, let $\Gamma_0$ be a cubic lattice on $\Sigma$ and let $\Gamma_n$ be $\Gamma_0$ subdivided $n$-times. In each subdivision every edge is split in two and new edges and vertices are introduced in such a way that the new lattice is again cubic, see Fig.~\ref{ronnie}. These graphs give rise to a coordinate system in such a way that the edges correspond to one unit of the coordinate axes and that the orientations of the coordinate axes coincide with the orientations of $\Sigma$.  This means, as already
mentioned in the introduction, that the full set of nested cubical lattices will
form what amounts to a coordinate system. Note, however, that~$\Sigma$ is not
equipped with a background metric.

Def\/ine f\/irst
\[
L^2(\ca_{\Gamma_n})=L^2\big(G^{e(\Gamma_n)}\big),
\]
where the measure on the right hand side is the normalized Haar measure.
Next def\/ine
\[
L^2(\overline{\ca})=\lim_nL^2 (\ca_{\Gamma_n}).
\]
This will be the Hilbert space on which we will def\/ine the quantized operators.
A path~$p$ in~$\cup_n \Gamma_n$  gives rise to a bounded function~$h_p$ with values in~$M_2$ via
\[
\overline{\ca}\ni \nabla \to \nabla(p),
\]
where $\nabla(p)$ is the extension of the holonomy map from $\ca$ to~$\overline{\ca}$, see e.g.~\cite{AGN3}. Therefore~$h_p$ has a natural action on the Hilbert space~$L^2(\overline{\ca})\otimes M_2$.

To construct the f\/lux operators f\/irst look at an edge $l\in \Gamma_n $. The f\/irst guess for a f\/lux operator associated to the inf\/initesimal surface $S_l$ sitting at the right end point of $l$ is
\[
\hat{E}^{S_l}_a=i \mathcal{L}_{e^a} ,
\]
where $\mathcal{L}_{e^a}$ is the left invariant vector f\/ield on the copy of $SU(2)$ associated to~$l$ corresponding to the generator~$e^a$ in $\mathfrak{su}(2)$ with index~$a$.
This guess is motivated by the bracket between $ \mathcal{L}_{e^a} $ and an element in~$SU(2)$
\[
[ \mathcal{L}_{e^a} , g] = g \sigma^a,
\]
which has the same structure as the Poisson bracket~(\ref{poisson1}) between classical f\/lux and loop variables. A careful analysis shows, however, that the correct formula reads
 \begin{gather*}
 \hat{E}^{S_l}_a=i \mathcal{L}_a+\mathcal{O}_{n-1}  ,
 \end{gather*}
where $\mathcal{O}_{n-1}$ is an operator which can be ignored in the particular limit with which we are concerned in the following\footnote{The appearance of $\mathcal{O}_{n-1}$ is related to the choice of projective system (\ref{subsub}), see below.}
(for details see~\cite{AGNP1}).

\subsection{A semif\/inite spectral triple of holonomy loops}\label{section5.2}

A spectral triple $(B,H,D)$ consists of three elements: a $\star$-algebra $B$ represented as bounded operators on a separable Hilbert space $H$ on which also an unbounded, self-adjoint operator $D$, called the Dirac operator, acts. The triple is required to satisfy the following two conditions:
\begin{itemize}\itemsep=0pt
\item
The resolvent of $D$, $(1+D^2)^{-1}$, is a compact operator in $H$.
\item
The commutator $[D,b]$ is bounded $\forall \, b\in B$.
\end{itemize}
The aim is now to build a spectral triple by rearranging the holonomy and f\/lux operators introduced in the previous sections. For details we refer to~\cite{AGN3,AGN2}. The triple consists of:
\begin{enumerate}\itemsep=0pt
\item
 the algebra generated by based holonomy loops running in the union $\cup_n\{\G_n\}$.
 \item
A Dirac operator which has the form
\begin{gather}
D=\sum_{i,a}  a_{n(i)}   {\bf e}_i^a \mathcal{L}_{e_i^a} .
\label{DDDD}
\end{gather}
 \end{enumerate}
Here, the algebra loops are again represented via matrix multiplication on the matrix factor in the Hilbert space
 \begin{gather}
 H=L^2\big(\overline{\ca}, Cl(T^*\overline{\ca})\otimes M_2(\mathbb{C})\big),
\label{HaHa}
 \end{gather}
 where we have introduced the complexif\/ied Clif\/ford bundle $Cl(T^*\overline{\ca})$ in order to accommodate the Dirac operator. For information on the construction of this Clif\/ford bundle see next paragraph. In principle we could here also take any $Cl(T^*\overline{\ca})$ module. This issue has so far not been considered.
This Hilbert space should again be understood as an inductive limit over Hilbert spaces associated to f\/inite lattices.
Also, $\mathcal{L}_{e_i^a}$ are the left-invariant vector f\/ields associated to the $i$'th copy of $G$ and where ${\bf e}_i^a$ is the associated element in the Clif\/ford algebra. The real constants~$a_{n(i)}$ represent a weight associated to the depth in the projective system of lattices at which the particular copy of $G$ appears.

The main technical tool introduced in \cite{AGN3,AGN2} in order to construct the Dirac operator is that the projective system of intermediate spaces~(\ref{intermediate}) can be rewritten into a system of the form
\begin{gather}
G^{e(\Gamma_0)}\leftarrow G^{e(\Gamma_1)}\leftarrow \cdots ,
\label{subsub}
\end{gather}
where the maps consist in deleting copies of $G$'s. Once this has been done the construction of the Dirac type operator (\ref{DDDD}) follows immediately as an inf\/inite sum over all the copies of $G$'s. This means that the sum in (\ref{DDDD}) runs over copies of $G$ which are assigned to the graphs in a very particular way.
The construction of the cotangent bundle $T^*\overline{\ca}$ of $\overline{\ca}$ is the associated system of cotangent bundles of $G^{e(\G_n)}$. To construct the complexif\/ied Clif\/ford bundle over $T^*\overline{\ca}$ one needs to choose a metric on $T^*\overline{\ca}$. This is done by f\/irst choosing a left and right invariant metric on $G$. The extension to a metric on each~$G^{e(\G_n)}$ compatible with the projective system~(\ref{subsub}) is straight forward. However, this choice of metric heavily depends on this particular choice of projective system (\ref{subsub}) (see~\cite{AGN2} for another possible choice). For details we refer to~\cite{AGN3}.
For details on the construction of the Dirac type operator we refer to~\cite{AGN3, AGN2}. In the appendix in~\cite{AGN2} the Dirac type operator (\ref{DDDD}) is written explicitly for one copy of $SU(2)$.

In \cite{AGN3} (see also \cite{Lai:2010ig}) it was proven that for $G=SU(2)$ this triple is a~semif\/inite spectral triple whenever the sequence $\{a_n\}$ approaches inf\/inite with $n$. The term `semif\/inite' refers to the fact that a residual symmetry group related to the Clif\/ford algebra acts in the Hilbert space. The resolvent of $D$ is therefore only compact up to this symmetry group. For details on the construction of the spectral triple we refer to~\cite{AGN1} and~\cite{AGN3}. For the original def\/inition of a~semif\/inite spectral triple see~\cite{Carey}.

This spectral triple encodes the kinematics of quantum gravity:
the holonomy loops generate the algebra; the corresponding vector f\/ields are packed in the Dirac type operator and their interaction reproduces the structure of the Poisson bracket~(\ref{poisson1}).

There is a relation to the kinematical Hilbert space found in LQG. Essentially, the Hilbert space (\ref{HaHa}) can be thought of as a kinematical Hilbert space somewhere between the kinematical and dif\/feomorphism invariant Hilbert space of LQG. For details see \cite{AGN3}.

\section{Semiclassical states}
\label{section6}

With the construction of the spectral triple in place the question arises what concrete physical structures can be derived from such a construction. A f\/irst attempt to address this question was made in a series of papers \cite{Aastrup:2011dt,Aastrup:2009dy,Aastrup:2010kb,Aastrup:2010ds, AGNP1} where it was shown that a natural class of semiclassical states resides within the Hilbert space $H$ and that the expectation value of $D$ on these states, in a semiclassical approximation combined with a certain continuum limit, entails an inf\/inite system of fermions interacting with gravity. Furthermore, in the particular limit where the semiclassical approximation is centered around a f\/lat space-time geometry, a fermionic Fock space emerges, along with a free fermionic quantum f\/ield theory. These results suggest that the spectral triple bridges between canonical quantum gravity and (fermionic) quantum f\/ield theory.

In this section we review these results. The f\/irst step is to introduce coherent states from which we subsequently construct the semiclassical states.

\subsection[Coherent states in $H$]{Coherent states in $\boldsymbol{H}$}\label{section6.1}

We start by recalling results for coherent states on various copies of $SU(2)$. This construction uses results of Hall \cite{H2, H1} and is inspired by the articles~\cite{BT1,BT2,TW}.

First pick a point $(A_n^a,E^m_b)$ in the phase space~$\cp$ of Ashtekar variables. The states which we construct will be coherent states peaked over this point.
Consider f\/irst a single edge~$l_i$ and thus one copy of~$SU(2)$. Let $\{{\bf e}^a_i\}$ be a basis for~$\mathfrak{su}(2)$.
There exist families $\phi^t_{l_i}\in L^2(SU(2))$ such that
\[
 \lim_{t \to 0}\langle \phi^t_{l_i}, t \cl_{{\bf e}^a_i}\phi_{l_i}^t \rangle=2^{-2n}\mathrm{i}E_a^m(x_{j}),
 \]
and
\[
\lim_{t \to 0}\langle \phi_{l_i}^t\otimes v, \nabla(l_i)\phi_{l_i}^t\otimes v \rangle=(v,h_{l_i}(A)v) ,
\]
where $v \in \bbC^2$, and $(,)$ denotes the inner product hereon; $x_{j}$ denotes the ``right'' endpoint of $l_i$ (we assume that $l_i$ is oriented to the ``right''), and the index `$m$' in the $E^m_a$ refers to the direction of $l_i$.
The factor $2^{-2n}$ is due to the fact that $L_{{\bf e}_j^a}$ corresponds to a f\/lux operator with a surface determined by the lattice \cite{AGNP1}.
Corresponding statements hold for operators of the type
\[
f(\nabla(l_i))P(t \cl_{{\bf e}^1_i},t \cl_{{\bf e}^2_i},t \cl_{{\bf e}^3_i}),
\]
where $P$ is a polynomial in three variables, and $f$ is a smooth function on $SU(2)$, i.e.
\[
 \lim_{t \to 0}\langle \phi^t_{l_i} f(\nabla(l_i))P(t \cl_{{\bf e}^1_i},t \cl_{{\bf e}^2_i},t \cl_{{\bf e}^3_i}) \phi^t_{l_i} \rangle=f(h_{l_i}(A))P(\mathrm{i}E_1^m,\mathrm{i}E_2^m,\mathrm{i}E_3^m) .
\]

Let us now consider the graph $\G_n$. We split the edges into $\{ l_i \}$ and $\{ l'_i\}$, where $\{ l_i\}$ denotes the edges appearing in the $n$'th subdivision but not in the $(n-1)$'th subdivision, and $\{ l'_i \}$ the rest. Let
$\phi^t_{l_i}$ be the coherent state on $SU(2)$ def\/ined above and
def\/ine the states $\phi_{l'_i}$ by
\[
\lim_{t \to 0}\langle \phi^t_{l'_i} \otimes v ,\nabla(l_i) \phi^t_{l'_i}\otimes  v \rangle =  (v,h_{l'_i}(A)v) ,
\]
and
\[
\lim_{t \to 0}\langle \phi^t_{l'_i} , t \cl_{{\bf e}_j^a} \phi^t_{l'_i}\rangle = 0 .
\]
Finally def\/ine $\phi^t_n$ to be the product of all these states as a state in $L^2(\ca_{\Gamma_n})$. These states are essentially identical to the states constructed in \cite{TW} except that they are based on cubic lattices and a particular mode of subdivision.

In the limit $n\rightarrow\infty$ these states produce the right expectation value on all loop operators in the inf\/inite lattice. The reason for the split of edges in $l_i$ and $l'_i$ in the def\/inition of the coherent state is to pick up only those degrees of freedom which 'live' in the continuum limit $n\rightarrow\infty$. In this way we shall, once the continuum limit is taken, partially have eliminated dependencies on f\/inite parts of the lattices. In a classical setup, this amounts to information which has measure zero in a Riemann integral.

\subsection{Semiclassical states and emergent matter}\label{section6.2}

The expectation value of $D$ on coherent states $\phi^t_n$ is zero since the Dirac operator is odd with respect to the Clif\/ford algebra and the coherent states do not take value therein. To f\/ind semiclassical states on which $D$ has a nonvanishing expectation value we introduce a generalized parallel transport operator ${\bf U}_p$.

Consider f\/irst the graph $\G_n$ and an edge $l_i$ in $\G_n$. Associate to $l_i$ the operator
\[
{\bf U}_i:=  \frac{\mathrm{i} }{2}\big(   {\bf e}_i^a g_i \sigma^a + \mathrm{i} {\bf e}_i^1{\bf e}_i^2{\bf e}_i^3 g_i\big)
\]
and check that ${\bf U}_i^*{\bf U}_i={\bf U}_i {\bf U}_i^*=\mathds{1}_2$. Here, $g_i=\nabla(l_i)$ is an element in the copy of $G$ assigned to the edge $l_i$.
Next, let $p=\{ l_{i_1}, l_{i_2},\ldots , l_{i_k}  \}$ be a path in $\G_n$ and def\/ine the associated operators by
\begin{gather*}
{\bf U}_p := {\bf U}_{i_1}{\bf U}_{i_2} \cdots {\bf U}_{i_k}   ,\qquad {U}_p := \nabla(l_{i_1})\cdot \nabla(l_{i_2})\cdots \nabla(l_{i_k}) ,
\end{gather*}
where $U_p$ is the ordinary parallel transport along $p$. The operators ${\bf U}_p$ form a family of mutually orthogonal operators labelled by paths in $\G_n$ such that
\[
 \mbox{Tr}_{\rm Cl}\left(  {\bf U}_p^* {\bf U}_{p'}\right) = \d_{p,p'} .
\]
This relation is due to the presence of the Clif\/ford algebra elements in ${\bf U}_p$. Here $\d_{p,p'}$ equals one when the paths $p$ and $p'$ are identical and zero otherwise.

Notice that if it were not for the second term in the def\/inition of ${\bf U}_i$ this operator would, up to a factor, equal a commutator
\[
[D,\nabla(l_i)] .
 \]
 In the language of noncommutative geometry such a commutator corresponds to a one-form \cite{ConnesBook} and therefore this suggests that ${\bf U}_i$ has a geometrical origin.

Consider now states in $\ch_{\G_n}$ of the form
\begin{gather}
\Psi^t_{m,n}(\psi_1,\ldots,\psi_m,\phi^t_n):= \sum_{p\in P_m} (-1)^p \Psi_{n}(\psi_{p(1)})\cdots\Psi_{n}(\psi_{p(m)})\phi_n^t,
\label{supermanN}
\end{gather}
where
\begin{gather}
\Psi_{n}(\psi) = 2^{-3n}\sum_i {\bf U}_{p_i}\psi(x_i) U^{-1}_{p_i} ,
\label{sss}
\end{gather}
where the sum runs over vertices $x_i$ in $\G_{n}$ and where the path $p_i$ connects the basepoint~$x_0$ with vertices $x_i$. Here $\psi(x_i)$ denotes an element in $M_2(\mathbb{C})$ associated to the vertex~$x_i$. The matrices~$\psi(x_i)$ will be seen to form a spinor degree of freedom at the point $x_i$ once the semiclassical limit is performed.

Note in passing that the expectation value of a loop operator $h_L$ on states involving ${\bf U}_p$'s amounts to taking a trace of the loop operator. This is due to the relation
\[
\langle  {\bf U}_i   g_i    {\bf U}_i  \rangle_{Cl} = \langle g_i \rangle_{M_2},
\]
where the l.h.s.\ involves the trace over the Clif\/ford algebra and the r.h.s.\ involves the trace over the matrix factor in~$H$. Since the base point dependency is absent for traced loops this means that the base point dependency vanishes on states of the form~(\ref{sss}).

It was shown in \cite{Aastrup:2010ds} that the expectation value of $D$ on the states (\ref{supermanN}) with $m=1$ gives a~spatial Dirac operator acting on a spinor
\begin{gather}
\lim_{n\rightarrow\infty}\lim_{t\rightarrow 0} \langle \Psi^t_{1,n}\vert t D \vert \Psi^t_{1,n}\rangle =
  \frac{1}{2}  \int_\Sigma d^3x  {\psi}^*(x) \big( E^m_a \nabla_m \sigma^a  + \nabla_m E^m_a  \sigma^a    \big) \psi(x)
 \nonumber\\
\hphantom{\lim_{n\rightarrow\infty}\lim_{t\rightarrow 0} \langle \Psi^t_{1,n}\vert t D \vert \Psi^t_{1,n}\rangle}{}
 =  \int_\Sigma d^3x  {\psi}^*(x) \notD \psi(x)  ,
 \label{hmmm}
\end{gather}
where $\nabla_m= \pa_m + A_m$, and where we f\/ixed the free parameters~$a_{n}$ in~$D$ to~$2^{3n}$.
The result~(\ref{hmmm}) shows that the role of the cubic lattices is that of the coordinate system in which we have written the Ashtekar variables. Furthermore, the emergence of the integral depends crucially on the presence of the Clif\/ford algebra elements in~$\Psi^t_{1,n}$.

To obtain instead the Dirac Hamiltonian, which involves also the lapse and shift variables encoding the foliation of space-time, one can modify the construction in dif\/ferent ways. For instance, one may add a certain class of matrix factors to the Dirac operator, see~\cite{Aastrup:2009dy}. It is, however, at the moment not clear which strategy is the right one.

Next, it was shown in~\cite{Aastrup:2011dt} that the expectation value of~$D$, with the same double limit, on the states $\Psi^t_{m,n}$ gives a system of~$m$ fermions coupled to the gravitational f\/ields around which the semiclassical approximation is performed
\begin{gather*}
\lim_{n\rightarrow\infty}\lim_{t\rightarrow 0} \langle \Psi^t_{m,n}\vert t D \vert \Psi^t_{m,n}\rangle =
\sum_{i=1}^m \int_\Sigma d^3x  {\psi}_i^*(x) \notD \psi_i(x) + \mbox{interaction terms} ,
\end{gather*}
where the interaction terms are of the form
\[
\int_\Sigma dx \int_\Sigma dy
 \mbox{Tr}\big( U(y,x) \psi^*_2(x) {\not\hspace{-1mm}\nabla} \psi_1(x) U(x,y) \psi^*_1(y)\psi_2(y)\big),
\]
where $U(x,y)$ are parallel transports connecting $x$ and $y$. The computations are rather elaborate and we refer to~\cite{Aastrup:2011dt} for details.

Thus, what emerges from the states (\ref{sss}) is an inf\/inite system of fermions coupled to gravity. These fermions come with additional interaction terms which involve f\/lux tubes of the Ashtekar connection between fermions at dif\/ferent locations in~$\Sigma$. In~\cite{Aastrup:2011dt}, however, another class of semiclassical states were also identif\/ied which did not entail this nonlocal interaction. Thus, the credibility of such an interaction is still unsettled. In any case, in the particular limit where the coherent states are peaked around the f\/lat space-time geometry, the whole system coincides with a system of free fermions. Thus, a free fermionic quantum f\/ield theory emerges.

\section{Discussion and outlook}\label{section7}

The usefulness of the intersection of noncommutative geometry and canonical quantum gravity, which has been presented in the previous sections, is demonstrated in two main results:
\begin{enumerate}\itemsep=0pt
\item
The spectral triple encodes the kinematics of quantum gravity. The interaction between the algebra of holonomy loops and the Dirac type operator, which is build from f\/lux operators, reproduces the Poisson structure of the corresponding classical variables. This result is obtained with manifestly separable structures.
\item
The spectral triple establishes a concrete link between canonical quantum gravity and fermionic quantum f\/ield theory. A natural class of semiclassical states entail, in a semiclassical approximation, an inf\/inite system of fermions and, ultimately, a Fock space structure.
\end{enumerate}
Thus, the overall picture emerges that it is possible to {\it derive} central elements of quantum f\/ield theory from a construction which a priori is only concerned with gravitational degrees of freedom.
These results raise a number of questions which we will elaborate on in the following.

Before we do that, let us emphasize that up to now we are far from having exploited the full range of the toolbox of noncommutative geometry to the framework of canonical quantum gravity. We believe that the intersection between the two deserves more analysis to establish its importance.

\subsection{The continuum limit}\label{section7.1}

The link to fermionic quantum f\/ield theory presented in Section~\ref{section6} arise from a double limit, where {\it first} a semiclassical approximation is taken from which classical gravitational variables emerge, and {\it second} a continuum limit is taken. Thus, what actually happens is that classical variables are inserted in a f\/inite lattice and subsequently a continuum limit is taken much alike the continuum limit in a Riemann integral. This procedure immediately raises the question whether it is possible to take the continuum limit alone without the semiclassical approximation.

Another reason for searching for a continuum limit of the construction is that the result that the space $\ca$ is densely embedded in the projective limit $\overline{\ca}$ does not depend on any f\/inite number of graphs. Thus, we can remove a f\/inite number of graphs and still separate $\ca$. This amounts to removing a set which classically has measure zero.
In fact the very presence of f\/inite graphs discretizes the underlying space.  For a sequence of paths $\{ p_n \}$ converging to $p$, the sequence of functions $\{ h_{p_n} \}$ will not converges in the norm (\ref{norm}) to $h_p$. This has the consequence that in $\overline{\ca}$ there are objects which are, for example, localized on a single edge, and therefore violates the topology and smooth structure of the underlying manifold. In particular def\/ining the curvature of such an object seems unnatural.  A result by Ashtekar and Lewandowski \cite{Ashtekar:1993wf} states that the spectrum of the algebra of Wilson loops is $\overline{\ca} / \overline{\cg}$, where $\overline{\cg}$ denote the group of generalized gauge transformations. This shows that building a theory based on single loops or paths as bounded observables automatically leads to a certain  discretization of the underlying manifold.

 Thus, it seems plausible that one should really only work with the inf\/initely ref\/ined graph, and that the f\/inite graphs should only play the role of auxiliary structures necessary to implement a~continuum limit.

The structure of the semiclassical states (\ref{sss}) suggests that a continuum limit is found by considering sequences of objects -- algebra elements and states -- assigned to each level in the projective system of graphs. The semiclassical states (\ref{sss}) come in this form
\begin{gather}
(\Psi_0,\Psi_1,\Psi_2,\ldots, \Psi_n, \ldots)\qquad\mbox{with}\qquad\Psi_n \in H_n =L^2(\ca_{\G_n},Cl(T^*G^n))\otimes M_2
\label{sequences}
\end{gather}
and the question arises what special requirements govern these sequences. In the case of the semiclassical states there seems to be a similarity map $\Psi_n\rightarrow\Psi_{n+1}$ which ensures that the states, at each level, has the right structure.
Another feature of the semiclassical states (\ref{sss}) is that they come with spinor degrees of freedom which give weight to loops located at dif\/ferent spatial points. Thus, these states come with information of the point set topology of an underlying manifold. Therefore, one might speculate whether a notion of local smooth structure should be introduced as a condition which regulates which sequences of the form (\ref{sequences}) one should permit.

These considerations lead us to propose to construct an algebra generated by inf\/inite sequences, where each entry is an algebra element assigned to a f\/inite graph and where certain scaling conditions link the dif\/ferent level to each other. These scaling conditions should be local in the sense of an underlying manifold, and they should, at least in a semiclassical approximation, entail smooth structures.

\subsection{An algebra of parallel transports}\label{section7.2}

In Section \ref{section4} we pointed out that there are three dif\/ferent ways in which one can build a $C^*$-algebra generated by parallel transports: one generated by Wilson loops, one by holonomy loops and one by open parallel transports. These three choices come with increasing noncommutativity, and since the key message coming from noncommutative geometry is that any noncommutative structure carries important information, it seems natural to consider the last option, an algebra of open-ended parallel transports.

Such an algebra is based on a groupoid structure where two elements have a non-zero product whenever their end- and start-points coincide. Thus, such an algebra will be generated by elements of the form
\[
\sum_i a_i h_{p_i},\qquad a_i\in\mathbb{C},
\]
which, at the $n$'th level of subdivision, can be represented on a Hilbert space which carries also information of the set of points in $\G_n$
\[
H'_n = H_n \times \{v_n\} ,   \qquad v_n \in\G_n,
\]
which means that one is to consider matrices of the size of the number of points in~$\G_n$ where $h_p$ is an entry~$(s(p),e(p))$ in such a matrix. The diagonal carries the loops and the trace in~$H'_n$ gives, in the limit $n\rightarrow\infty$, an integral over gauge invariant objects.

Notice that the parallel transports are partial isometries and that the spectrum of such an algebra is not expected to be larger than the equivalent algebra generated by loops since the irreducible representations will be the same (possible up to gauge equivalence). Thus, this is a~standard example of a quotient taken in the noncommutative manner as described in Section~\ref{section2}. Here we have used the parallel transports to noncommutatively identify the dif\/ferent base points. Such an algebra can therefore be seen as the noncommutative version of the algebra of Wilson loops.

Let us consider what inner f\/luctuations of the Dirac type operator with elements in such an algebra, denoted $ \cb_{\rm  parallel}$, will give:
\begin{gather}
D\rightarrow \tilde{D} := D + a[D,b], \qquad a,b\in \cb_{\rm  parallel}.
\label{inner}
\end{gather}
Let us also assume that we have implemented a continuum limit, as proposed in the previous section, which means that the parallel transports in $a$ and $b$ are inf\/initesimal in the sense of the lattices. The commutator of $D$ with a single parallel transport $\nabla(l_i)$ along an edge $l_i$ is simply
\[
[D,\nabla(l_i)] =  \mathrm{i} a_{n(i)}  {\bf e}_i^a \nabla(l_i) \sigma^a
\]
and when this is inserted into the computations leading to (\ref{hmmm}) we f\/ind that the spatial Dirac operator is modif\/ied to
\[
 \notD \rightarrow  \notD +  \notA,
\]
where $A=A(x)$ is a f\/ield emerging from the weights in $a$ and $b$. Although this computation is not strictly rigorous, due to the lack of precise def\/initions etc., it does suggest that what appears to be a gauge f\/ield, which is not a spin connection, emerge in a semiclassical approximation from f\/luctuations of $D$ of the kind~(\ref{inner}).

Note also that since the Ashtekar variables, and with them the loop and f\/lux variables used in loop quantum gravity, involve the spin-connection, it is natural that a construction based on them should involve fermions. A parallel transport moves a fermion from one point to another. It does, however, not appear natural to consider single paths, since an operation herewith is singular in the sense that it does not preserve smooth structures of the underlying manifold, see the discussion in the subsection on the continuum limit. Rather, it seems natural to consider a smooth collection of parallel transports which moves a (local) fermion f\/ield from one open set to another. Again, this points towards an algebra obtained in a continuum limit which is
 generated by operators of parallel transports with support in the inf\/initely ref\/ined lattice and which transport a fermion f\/ield from one open set to another in a smooth manner.

\subsection{The question about dif\/feomorphisms}\label{section7.3}

The introduction of a countable system of graphs immediately raises the question whether one can reconcile this choice with an action of the dif\/feomorphism group. Basically, the motivation for working with the uncountable set of piece-wise analytic graphs in loop quantum gravity is exactly that this ensures an action of  (analytic) dif\/feomorphisms which simply move the graphs around. Such an action is clearly absent in a construction which involves only loops running in an inf\/inite system of cubic lattices.

On the other hand, the expression which emerges in (\ref{hmmm}), as well as the emergence of the Dirac Hamiltonian from a modif\/ication of the Dirac type operator, {\it is} manifestly invariant. Thus, it seems that {\it if} an action of the dif\/feomorphism group can be introduced in a construction based on cubic lattices, then it should happen in the continuum limit proposed above. This is the limit where the states which entail (\ref{hmmm}) live, and it is the limit where the interpretation of the lattices as a Cartesian coordinate system becomes apparent.

Dif\/feomorphism invariance is encoded in the algebra of the constraints (\ref{constraints}). An unbroken algebra of quantized constraint operators is equivalent to having an unbroken action of the dif\/feomorphism group. Thus, to check whether a construction based on cubic lattices is reconcilable with dif\/feomorphisms one needs to def\/ine quantized constraint operators and compute their algebra. Preliminary, unpublished results show that for instance the bracket between two quantized Hamilton operators, $\hat{H}(N)$, $\hat{H}(N')$ where $N$ and $N'$ are dif\/ferent lapse f\/ields, give
\[
[\hat{H}(N),\hat{H}(N')] = \int d^3x (N \pa^m N' - N' \pa^m N) \hat{D}_m + \OO,
\]
where $\hat{D}_m$ is the quantized dif\/feomorphism constraint. $\OO$ is an anomalous term which involves factors
\begin{gather}
\big( \hat{E}^{S_{l_i}}_a -  \hat{E}^{S_{l_{i+1}}}_a\big),
\label{factor}
\end{gather}
where $l_i$ and $l_{i+1}$ are two parallel edges one lattice spacing apart. In a continuum limit as proposed above this lattice spacing will approach zero and the factor (\ref{factor}) becomes a measure of continuity of the (expectation value) of the triad operators  $\hat{E}^{S_l}_a$. Thus, with a subtle def\/ined continuum limit we believe that this bracket can close on a domain given by a semiclassical approximation. This should be understood in the sense that the bracket will close to any order in a given semiclassical approximation~-- but will be anomalous on the full domain of the Hilbert space. This ref\/lects, we believe, the viewpoint that dif\/feomorphisms are ill def\/ined outside a~domain with a notion of a smooth manifold, for instance in a state at very high temperature.

\subsection[Tomita-Takesaki and dynamics]{Tomita--Takesaki and dynamics}\label{section7.4}

The question about dif\/feomorphisms is of course closely related to the question about how a~dynamical principle is introduced. One approach is to simply write down an operator which quantizes the classical Hamiltonian and then check that these coincide in a classical limit. This is the standard quantization approach which is also applied in loop quantum gravity (see for example~\cite{Thiemann:2007zz}). We believe that this approach is, ultimately, unsatisfactory for the following two reasons:
\begin{enumerate}\itemsep=0pt
\item
Such a quantization procedure always comes with multiple {\bf ambiguities}, such as ordering ambiguities, ambiguities concerning higher order corrections, as well as ambiguities about various choices made in the quantization procedure.
\item
On a similar note, such a procedure is unlikely to uncover a possible deep mathematical mechanism that govern the dynamics of a unif\/ied theory of quantum gravity. Put dif\/ferently, it will not expose a fundamental {\bf principle} behind such a theory.
\end{enumerate}
As an alternative approach, we propose to apply the theory of Tomita and Takesaki to the general setup presented in this paper. Given a von Neumann algebra and a cyclic and separating state hereon, this theory identif\/ies a canonical, one-parameter group of automorphisms. In the present setup one might consider an algebra of parallel transports together with spectral projections of the Dirac operator. Natural states to consider are coherent states or semiclassical states (\ref{sss}), since, as already mentioned, these states entail dif\/ferentiable structure and thereby allow for dif\/feomorphisms. The Hamiltonian is then expected to be related to the modular operator.

To obtain a nontrivial application of Tomita--Takesaki theory the algebra is required to be highly noncommutative, preferably of type III. Thus, it seems likely that this idea is only reali\-zab\-le if one operates with a continuum limit of an algebra generated by parallel transports, as discussed in the previous paragraphs. The paper \cite{Kam} seems to support this statement, i.e.\ that the usual algebra of LQG is not suited for applying Tomita--Takesaki theory.

\subsection{A complex Ashtekar connection}\label{section7.5}

The original Ashtekar variables involve a complexif\/ied $SU(2)$ connection. Since the construction of the canonical Hilbert space in loop quantum gravity requires the gauge group to be compact~-- essentially because the identity function must be integrable in order for the inductive limit to be well def\/ined~-- the approach adopted in loop quantum gravity is to work instead with a real $SU(2)$ connection. This, however, means that the constraints no longer have the simple form in~(\ref{constraints}) and that the Ashtekar connection looses its geometrical interpretation \cite{Nicolai:1992xx}.

One might think that one can obtain a complex connection by simply doubling the Hilbert space corresponding to the real and complex parts, with the complex $\mathrm{i}$ given by the isometry between the two. This would work if we were working with the Lie-algebra, since the connection takes values herein, but the constructions based on algebras of parallel transports described in this paper operate with the Lie-groups, where
\[
g_i \sim {\rm Hol}(A,l_i)
\]
is the interpretation of the group elements $g_i$.

In the continuum limit which we have advocated above this will no longer hold true. The continuum limit is meant to single out the inf\/initely ref\/ined graph, where only edges which are inf\/initesimal with respect to the lattices appear. If we combine this with a semiclassical approximation via coherent states, then we would obtain
\[
g_i \sim \mathbf{1} +  A ds  + \co(ds^2),
\]
where $ds$ is now the lattice length of an inf\/initesimal edge~$l_i$. With such a relation one can indeed obtain a complexif\/ied connection by doubling the construction. This would, of course, only hold within the domain of the semiclassical approximation, but this is suf\/f\/icient since this is where classical gravity emerges.

The partial isometry which interchanges the two Hilbert spaces will here play the role of the complex~$\mathrm{i}$. Within the machinery of noncommutative geometry, it seems natural that this operator should be related to the real structure which plays a central role in noncommutative geometry.

\vspace{-1mm}

\subsection*{Acknowledgements}

We thank the referees for useful comments and suggestions.

\vspace{-2mm}

\pdfbookmark[1]{References}{ref}
\LastPageEnding

\end{document}